\documentclass[a4paper,12pt]{article}
\usepackage[top=25truemm,bottom=25truemm,left=15truemm,right=15truemm]{geometry}
\usepackage[]{hyperref}
\usepackage{amsmath}
\usepackage{ascmac}
\usepackage{amssymb}
\usepackage[]{graphicx} 
\graphicspath{{./figs/}}

\title{Interface in Kerr-AdS black hole spacetime}
\author{Koichi Nagasaki}
\date{\today}

\begin{document}
\vspace{1cm}
\begin{center}
{\LARGE Interface in Kerr-AdS black hole spacetime}\\
\vspace{2cm}
{\large Koichi Nagasaki}\footnote{koichi.nagasaki24@gmail.com}\\
\vspace{1cm}
{\small School of Physics,
University of Electronic Science and Technology of China,\\
Address: No.4, Section 2, North Jianshe Road, Chengdu 610054, China}
\end{center}
\vspace{1.5cm}
\abstract{
A defect solution in the AdS$_5\times S^5$ black hole spacetime is given.
This is a generalization of the previous work \cite{Nagasaki:2011ue} to another spacetime which includes rotating black holes.
The equation of motion for a sort of non-local operator, ``an interface," is given and we solve it by the numerical calculation.
This result gives a new example of non-equilibrium systems in the AdS/CFT including the gauge flux on the probe brane.
}
\vspace{1cm}

\tableofcontents

\section{Introduction}
Non-local operators are a useful tool for studying the AdS/CFT correspondence.
An interface is an example of this sort of operator realized in the D3/D5 brane system.
We studied this interface in the flat AdS$_5\times S^5$ spacetime in \cite{Nagasaki:2011ue}.
These brane systems are related to non-equilibrium systems in the AdS/CFT correspondence \cite{Das:2010yw}.
In the present work, we would like to consider the space times which has a structure AdS$_5\times S^5$ with  a gauge flux on the D5-brane.
In our past work \cite{Nagasaki:2011ue} we studied the interface solution in AdS spacetime AdS$_5\times S^5$ formed by D3-branes. 
This work showed that non-local operators introduce a new parameter which enables us to compare the results from a gauge theory and gravity theory, and then actually proved that the results from both theories agree in the first order. 
This parameter corresponds to the flux on the probe brane.
Introducing the flux is important also in the context of the compactification \cite{Koerber:2007jb}.
In this paper we treat an interface in black hole spacetime.
This analysis reveals the behavior of the interface in black hole spacetimes. 

Recently ``complexity" is studied to reveal black hole physics \cite{Susskind:2018pmk}.
Complexity-Action relation \cite{Brown:2015bva, Brown:2015lvg} is a very interesting conjecture.
This conjecture is recently studied in various spacetimes \cite{Brown:2018bms}.
According to this conjecture the quantity, holographic complexity, which describes the development of the inside of the black hole is equal to the action calculated in a region called the Wheeler-DeWitt patch.
This is a certain spacetime region defined by a given time at each boundary of the black hole spacetime.
This is a holographic relation. 
Then there should be a quantity which describes the counterpart in the boundary gauge theories.
By this motivation, recent works have tried to find the meaning of complexity in gauge theories \cite{Gan:2017qkz, Jefferson:2017sdb, Chapman:2018dem, Chapman:2018hou, Guo:2018kzl, Chapman:2018lsv, Khan:2018rzm, Brown:2019whu}.
Non-local properties are found in the context of complexity \cite{Fu:2018kcp}.
Then non-local operators studied in this paper are also related to this topic.

By these motivations we would like to study the more types of non-local operators. 
We studied the effect of the fundamental string for the complexity growth in the previous works \cite{Nagasaki:2017kqe, Nagasaki:2018csh}.
So we are next interested in more various dimensional nonlocal operators.
In this paper the goal is to find the embedding of the probe D5-brane which realizes a non-local operator, an interface, including a gauge flux.

The interface is introduced in gauge theories as a defect to separate the boundary space into two gauge theories and actively studied recently, for example, in \cite{Gutperle:2018fkz, Ovgun:2018jbm}.
In our previous work \cite{Nagasaki:2018dyz} we found that the interface can penetrate the horizon only for the case where the D5-brane has no gauge flux.
In this paper we study a different kind of the black hole which has the angular momentum. 
 
In our assumption the probe brane occupies the $S^3$ subspace of AdS$_5$. 
There are two ways for parametrising this space.
One is hyperspherical coordinates where the $S^3$ metric is written as
\begin{equation}\label{eq:metric_hyper}
ds^2 = d\vartheta^2 + \sin^2\vartheta(d\psi_1^2 + \sin^2\psi_1 d\psi_2^2).
\end{equation}
On the other hand we can choose other polar coordinates (Hopf coordinates) as
\begin{equation}\label{eq:hopfcoord}
x^1 = \sin\theta\sin\phi_1,\;
x^2 = \sin\theta\cos\phi_1,\;
x^3 = \cos\theta\sin\phi_2,\;
x^4 = \cos\theta\cos\phi_2,
\end{equation}
where the metric is expressed as
\begin{equation}\label{eq:metric_hopf}
ds^2 = d\theta^2 + \sin^2\theta d\phi_1^2 + \cos^2\theta d\phi_2^2.
\end{equation}
We used former coordinates system \eqref{eq:metric_hyper} in the previous analysis \cite{Nagasaki:2018dyz}.
If we take the former case (hyperspherical coordinates), the equation of motion is  
\begin{equation}\label{eq:EOMiAdSBH_hyper}
\frac{d}{dz}\Big(\frac{\varphi'z^2f(z)}{\sqrt{1+\varphi'^2z^2f(z)}}\Big)
= \frac4{z}\frac{\varphi'z^2f(z)}{\sqrt{1+\varphi'^2z^2f(z)}}
 - \frac{4K}{z}
 - \frac{2\tan\varphi}{\sqrt{1+\varphi'^2z^2f(z)}}.
\end{equation}
We use the latter coordinates system \eqref{eq:hopfcoord} in this paper.
This coordinate system includes two circles parametrized by $\phi_1$ and $\phi_2$.
The radii of these circles change depending on the latitude angle $\theta$.

This paper is organized as follows.
In section \ref{sec:iAdSSchBH_hopf} we consider the interface in AdS$_5\times S^5$ black hole spacetime without the angular momentum.  
We use the Hopf coordinates where the $S^3$ metric is written as \eqref{eq:metric_hopf}.
This black hole spacetime corresponds to the case where the angular momentum is zero in the Kerr-AdS black hole solution.
In section \ref{sec:iKABH} we introduce the angular momentum.
This is the Kerr-AdS black hole spacetime.
We investigate the gauge flux dependence of the interface solution.
The D5-brane does not penetrate the horizon for non zero gauge flux as expected from the previous analysis.
We also find the angular momentum dependence of the interface solution.

\section{Interfaces in the AdS-Schwarzschild black hole spacetime}\label{sec:iAdSSchBH_hopf}
In this section we find the interface solution in AdS$_5\times S^5$ black hole spacetime.
This is the solution in the same background to Ref.\cite{Nagasaki:2018dyz} with other coordinates and assumptions.
\subsection{Action and Equation of motion}
We set the coordinates as
\begin{align}
\text{AdS}_5&:\qquad
	t, \phi_1, \phi_2, r, \vartheta;
&& \vartheta\in[0,\pi/2],\qquad 
	\phi_1,\phi_2\in[0,2\pi),\nonumber\\
S^5&:\qquad
	\theta, \phi, \varphi_1, \varphi_2, \varphi_3;
&& \theta,\phi,\varphi_1,\varphi_2\in[0,\pi],\qquad
	\phi_3\in[0,2\pi).
\end{align}

The worldvolume of the D5-brane stretches in the subspace spanned by $(t,\phi_1,\phi_2,r,\vartheta,\theta,\phi)$.
It forms one-dimensional subspace in $(r,\vartheta)$ plane.
The compact part, $S^5$, is induced to
$
ds_{S^2}^2 = d\theta^2 + \sin^2\theta d\phi^2
$
on the D5-brane.
For the non-compact part, AdS$_5$, the bulk metric is
\begin{align}\label{eq:AdSBH_metric}
ds_\text{AdS$_5$}^2
&= -f(r)dt^2
 + r^2(\sin^2\vartheta d\phi_1^2 + \cos^2\vartheta d\phi_2^2)
 + \frac{dr^2}{f(r)}
 + r^2d\vartheta^2,\\
f(r) &= 1+r^2-\frac{r_\text{m}^2}{r^2}.\nonumber
\end{align}
In the above the parameter $r_\text{m}$ is related to the black hole mass through
\begin{equation}
r_\text{m}^2 
= \frac{16\pi GM}{5\text{vol}[S^3]}
= \frac{8GM}{5\pi}.
\end{equation}
In the $(r,\vartheta)$ plane, $\vartheta$ is a function of $r$:
\begin{equation}
\vartheta = \vartheta(r).
\end{equation}
This system has the SO(3) symmetry. 
Then, by the symmetry, this flux must be proportional to the volume form of $S^2$ subspace:
\begin{equation}
\mathcal F = -\kappa d\theta\wedge\sin\theta d\phi.
\end{equation}
There is the Ramond-Ramond 4-form defined as
\begin{equation}
C_4 = -r^4dt\wedge d\vartheta\wedge(\sin\vartheta d\phi_1)\wedge (\cos\vartheta d\phi_2)
\end{equation}
to satisfy $dC_4 = 4d\text{vol}[\text{AdS}_5]$.
From these ansatzes the induced metric is 
\begin{equation}\label{eq:dsindAdS5S5}
ds_\text{ind}^2
= -f(r)dt^2 + \Big(\frac{1}{f(r)}+r^2\vartheta'(r)^2\Big)dr^2 
  + r^2(\sin^2\vartheta d\phi_1^2 + \cos^2\vartheta d\phi_2^2)
  + d\theta^2 + \sin^2\theta d\phi^2.
\end{equation}
Taking the summation with the gauge flux, its determinant is
\begin{align}
&G_\text{ind} + \mathcal F = 
\begin{bmatrix}
-f(r) & & & \\
 & 1/f(r) + r^2\vartheta'^2& & & &\\
 & & r^2\sin^2\vartheta & & &\\
 & & & r^2\cos^2\vartheta& &\\
 & & & & 1& -\kappa\sin\theta\\
 & & & & \kappa\sin\theta& \sin^2\theta
\end{bmatrix},\nonumber\\
&\Rightarrow
\sqrt{-\det[G+\mathcal F]}
= \sqrt{1+\kappa^2}\sin\theta\;r^2\sin\vartheta\cos\vartheta\sqrt{1+\vartheta'^2r^2f(r)}.
\end{align}
The Wess-Zumino term is, from the gauge flux and the Ramond-Ramond 4-form,
\begin{equation}
S_\text{WZ}/T_5 = \int\mathcal F\wedge C_4
= (2\pi)^2\text{vol}[S^2]T\kappa\int dr\sin\vartheta\cos\vartheta r^4\vartheta'(r),
\end{equation}
where vol[$S^2$] is the volume spanned by $\theta$ and $\phi$ and $(2\pi)^2$ comes from the integral on the two circles spanned by $\phi_1$ and $\phi_2$.
The D5 brane action is
\begin{align}
S_\text{D5}/T_5 
&= -\int \sqrt{-\det[G_\text{ind}+\mathcal F]}
 + \int \mathcal F\wedge C_4,\nonumber\\
&= -(2\pi)^2\text{vol}[S^2]T\sqrt{1+\kappa^2}
  \int dr\sin\vartheta\cos\vartheta r^2(\sqrt{1+r^2f(r)\vartheta'(r)^2} - Kr^2\vartheta'(r)),
\end{align}
where we use $K := \kappa/\sqrt{1+\kappa^2}$ for later convenience.
This action is rewritten in coordinates $z=1/r$ and $\varphi=\pi/4-\vartheta$ as
\begin{align*}
\frac{S_\text{D5}}{T_5(2\pi)^2\text{vol}[S^2]T} 
&= -\sqrt{1+\kappa^2}
  \int dr\sin\vartheta\cos\vartheta r^2(\sqrt{1+r^2f(r)\vartheta'(r)^2} - Kr^2\vartheta'(r))\\
&= -\sqrt{1+\kappa^2}
  \int\frac{dz}{2z^4}\cos(2\varphi)(\sqrt{1+z^2f(z)\varphi'(z)^2} - K\varphi'(z))
\\
&= -\frac{\sqrt{1+\kappa^2}}2
  \int dz\frac{\cos(2\varphi)}{z^4}(\sqrt{1+g(z)\varphi'(z)^2} - K\varphi'(z)),
\end{align*}
where we defined a function,
$
g(z) := z^2f(z) = 1+z^2-r_\text{m}^2z^4.
$
From this action the equation of motion is 
\begin{equation}
\frac{d}{dz}\Big(\frac{\cos(2\varphi)}{z^4}\Big(\frac{g(z)\varphi'(z)}{\sqrt{1+g(z)\varphi'(z)^2}} - K\Big)\Big)
+ \frac{2\sin(2\varphi)}{z^4}(\sqrt{1+g(z)\varphi'(z)^2} - K\varphi'(z)) 
= 0.
\end{equation}
By operating the differentiation of $\cos(2\varphi)/z^4$, this equation is rewritten as follows.
\begin{equation}\label{eq:EOMiAdSBHhopf}
\frac{d}{dz}\Big(\frac{\varphi'g(z)}{\sqrt{1+\varphi'^2g(z)}}\Big)
= \frac4{z}\frac{\varphi'g(z)}{\sqrt{1+\varphi'^2g(z)}}
 - \frac{4K}{z}
 - \frac{2\tan(2\varphi)}{\sqrt{1+\varphi'^2g(z)}}.
\end{equation}
Eq.\ref{eq:EOMiAdSBHhopf} differs from the hyperspherical coordinates case (eq.\eqref{eq:EOMiAdSBH_hyper}) only by the factor two in the argument of the tangent function.

Introducing a parameter expression $(r(\tau),\varphi(\tau))$, the Lagrangian is
\begin{equation}
\frac{S_\text{D5}}{T_5(2\pi)^2\text{vol}[S^2]T} 
= -\int d\tau\mathcal L,\;\;
\mathcal L = \frac{\sqrt{1+\kappa^2}}2\frac{\cos(2\varphi)}{z^4}
(\sqrt{\dot z^2+g\dot\varphi^2} - K\dot\varphi).
\end{equation}
Since there is a gauge invariance we can choose the gauge where $\dot z^2+g\dot\varphi^2 = 1$.
The equations of motion are
\begin{subequations}
\begin{align}
z:\qquad&
\frac{d}{d\tau}\Big(\cos(2\varphi)\frac{\dot z}{z^4}\Big)
+ \frac{4\cos(2\varphi)}{z^5}(1 - K\dot\varphi)
- \frac{\cos(2\varphi)}{2z^4}g'\dot\varphi^2
= 0,\\
\varphi:\qquad&
\frac{d}{d\tau}\Big(
\frac{\cos(2\varphi)}{z^4}(g\dot\varphi - K)\Big)
+ \frac{2\sin(2\varphi)}{z^4}(1 - K\dot\varphi)
= 0.
\end{align}
\end{subequations}
Let us define $u_z := dz/d\tau, u_\varphi := d\varphi/d\tau$.
Then the equations are separated into the following four equations.
\begin{subequations}\label{eq:iAdSBH_Hopf_eom}
\begin{align}
\frac{dz}{d\tau} &= u_z,\\
\frac{d\varphi}{d\tau} &= u_\varphi,\\
\frac{du_z}{d\tau}
&= 2u_zu_\varphi\tan(2\varphi)
+ \frac4z(Ku_\varphi - gu_\varphi^2)
+ \frac{g'u_\varphi^2}2,\\
\frac{du_\varphi}{d\tau}
&= - \frac{g'}{g}u_zu_\varphi
+ 2\tan(2\varphi)(u_\varphi^2 - \frac1g) 
+ \frac{4u_z}{z}(u_\varphi - \frac{K}{g}).
\end{align}
\end{subequations}

\paragraph{Initial condition}
Since near the boundary $z\approx0$ the solution looks the same as the flat case \cite{Nagasaki:2011ue}, then $\varphi = \arcsin(\kappa z)$.
Taking account of the gauge condition, $u_z^2 + u_\varphi^2g = 1$, the initial condition is for small $z_0$
\begin{subequations}\label{eq:AdSBH_Hopf_initial}
\begin{align}
\varphi(z_0) 
&= \arcsin(\kappa z_0),\\
u_z(z_0) 
&= \sqrt\frac{1-\kappa^2z_0^2}{1+\kappa^2(1-r_\text{m}^2z_0^4)},\\
u_\varphi(z_0) 
&= \frac\kappa{\sqrt{1+\kappa^2(1-r_\text{m}^2z_0^4)}}.
\end{align}
\end{subequations}
Here we have to confirm the validity of the constraint.
Let us examine the evolution of the constraint.
By substituting the equations \eqref{eq:iAdSBH_Hopf_eom},
\begin{align}
&\frac{d}{d\tau}(u_z^2+gu_\varphi^2)
= 2u_z\dot u_z + 2gu_\varphi\dot u_\varphi + u_\varphi^2g'u_z\nonumber\\
&= 2u_z
\Big(2u_zu_\varphi\tan(2\varphi) 
 + \frac4z(Ku_\varphi + u_z^2 -1) + \frac{g'u_\varphi^2}2\Big)\nonumber\\
&\qquad
 + 2gu_\varphi
\Big(-\frac{g'}{g}u_zu_\varphi + 2\tan(2\varphi)(u_\varphi^2-\frac1g) 
 + \frac{4u_z}z(u_\varphi - \frac{K}g)\Big) 
 + u_\varphi^2g'u_z\nonumber\\
&= 4u_z^2u_\varphi\tan(2\varphi)
 + \frac{8u_zu_\varphi}{z}K + \frac{8u_z^3}{z} - \frac{8u_z}{z}
 + g'u_zu_\varphi^2\nonumber\\
&\qquad
 - 2g'u_zu_\varphi^2 + 4\tan(2\varphi)(gu_\varphi^3 - u_\varphi)
 + \frac{8gu_zu_\varphi^2}{z}
 - \frac{8u_zu_\varphi K}{z} + u_\varphi^2g'u_z\nonumber\\
&= 4u_\varphi(u_z^2+gu_\varphi^2-1)\tan(2\varphi)
 + \frac{8u_z}z(u_z^2+gu_\varphi^2-1)\nonumber\\
&= \Big(4u_\varphi\tan(2\varphi) + \frac{8u_z}z\Big)(u_z^2+gu_\varphi^2-1)
= 0.
\end{align}
where in the last equality we use the constraint 
$
u_z^2+gu_\varphi^2 = 1
$.
Then this constraint is valid.

\subsection{Interface solution}\label{subsec:iAdSBH_Hopf_sol}
By the numerical calculation, we solve the equations of motion \eqref{eq:iAdSBH_Hopf_eom} under the initial condition \eqref{eq:AdSBH_Hopf_initial}.
The result is shown in the following three figures.
The interface touches the boundary at the position $\varphi = 0$ and extends toward the inside of the AdS spacetime.
The steep at the beginning depends on the gauge flux.

\begin{figure}[h]
	\begin{minipage}[t]{0.5\linewidth}
	\includegraphics[width=\linewidth]{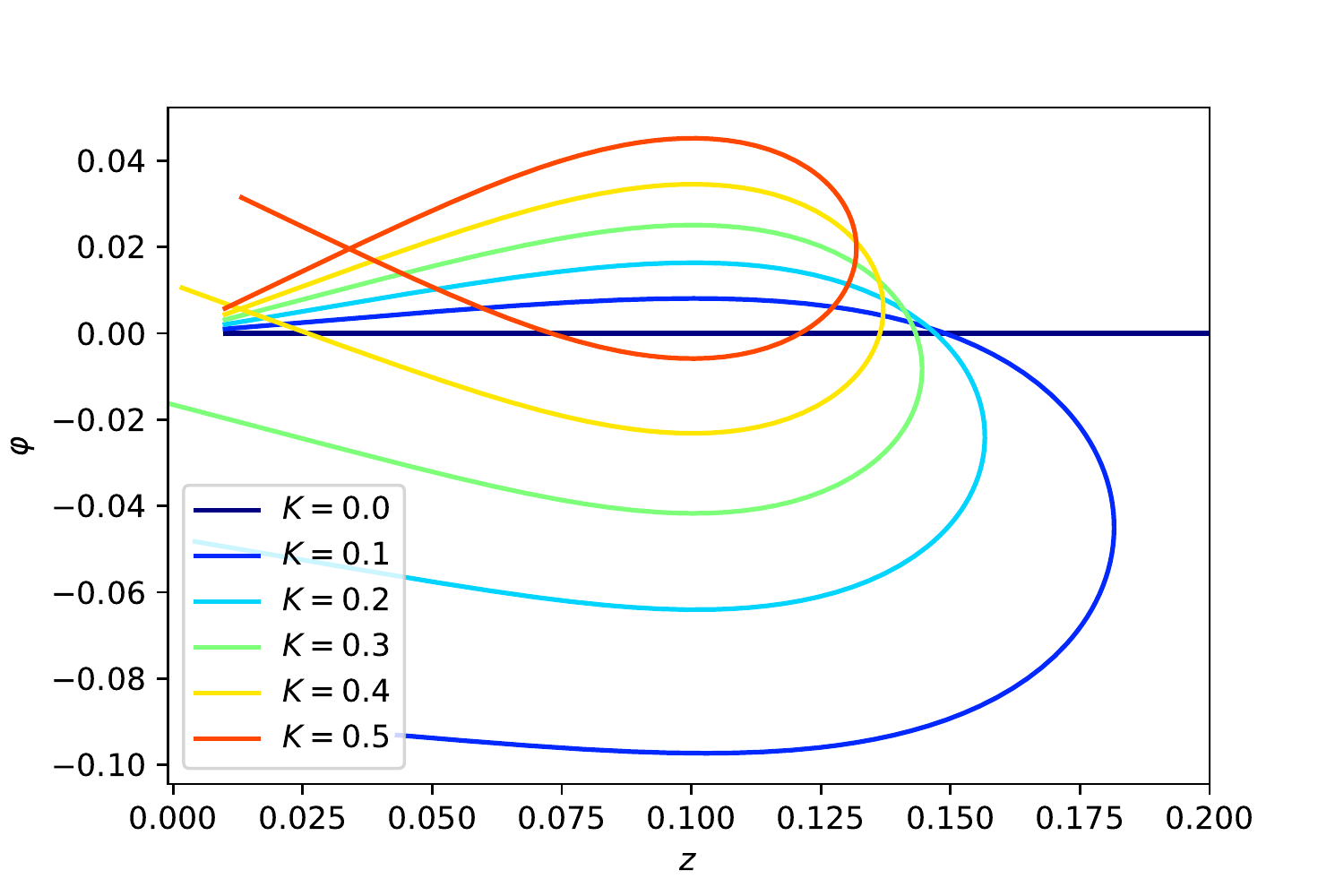}
	\caption{Flux dependence: mass $r_\text{m} = 10$}
	\label{fig:iAdSBH_hopf_K}
	\end{minipage}
\hspace{0.01\linewidth}
	\begin{minipage}[t]{0.5\linewidth}
	\includegraphics[width=\linewidth]{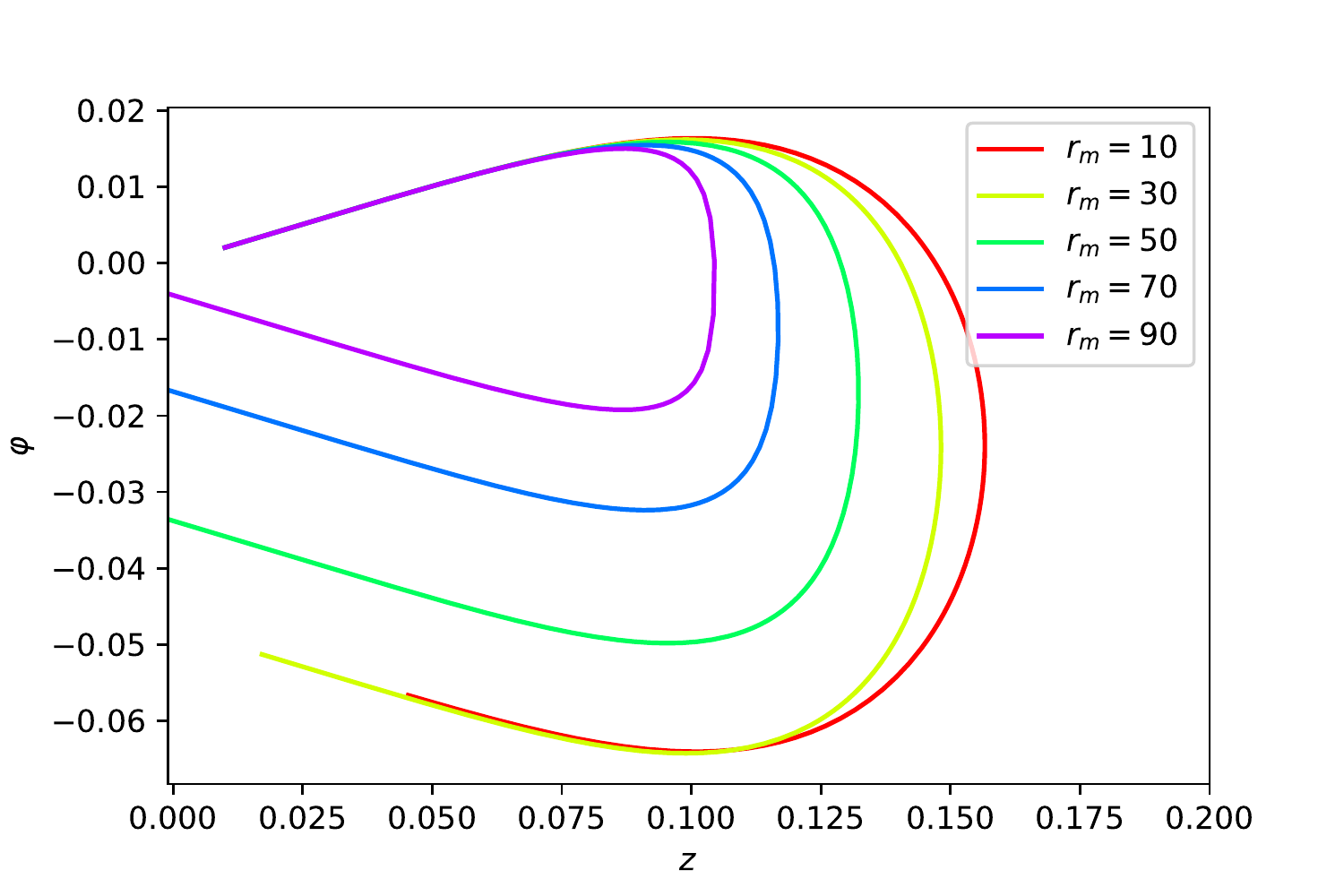}
	\caption{Mass dependence: $K = 0.2$}
	\label{fig:iAdSBH_hopf_M}
	\end{minipage}
\end{figure}

Figure \ref{fig:iAdSBH_hopf_K} shows the flux dependence of the D5-brane solution.
The black hole mass is fixed to $r_\text{m} = 10$.
The degree of the slope at the boundary changes depending on the strength of gauge flux. 
We can see that the D5-brane can penetrates the horizon only for no gauge flux case.

Figure \ref{fig:iAdSBH_hopf_M} shows the mass dependence for fixed gauge flux $K = 0.2$.
The D5-brane curves tightly for large masses to avoid the horizon.
Note that the position of the horizon in $z$ coordinates is given by 
\begin{equation}
z^2f(z) = 1+z^2-r_\text{m}^2z^4 = 0,\;
z_\text{h}^2 = \frac{1+\sqrt{1+4r_\text{m}^2}}{2r_\text{m}^2}.
\end{equation}
For examples, $z_\text{h} = 0.324$ for $r_\text{m} = 10$ and  $z_\text{h} = 0.105$ for $r_\text{m} = 90$.

\begin{figure}[h]
	\begin{minipage}[t]{0.5\linewidth}
	\includegraphics[width=\linewidth]{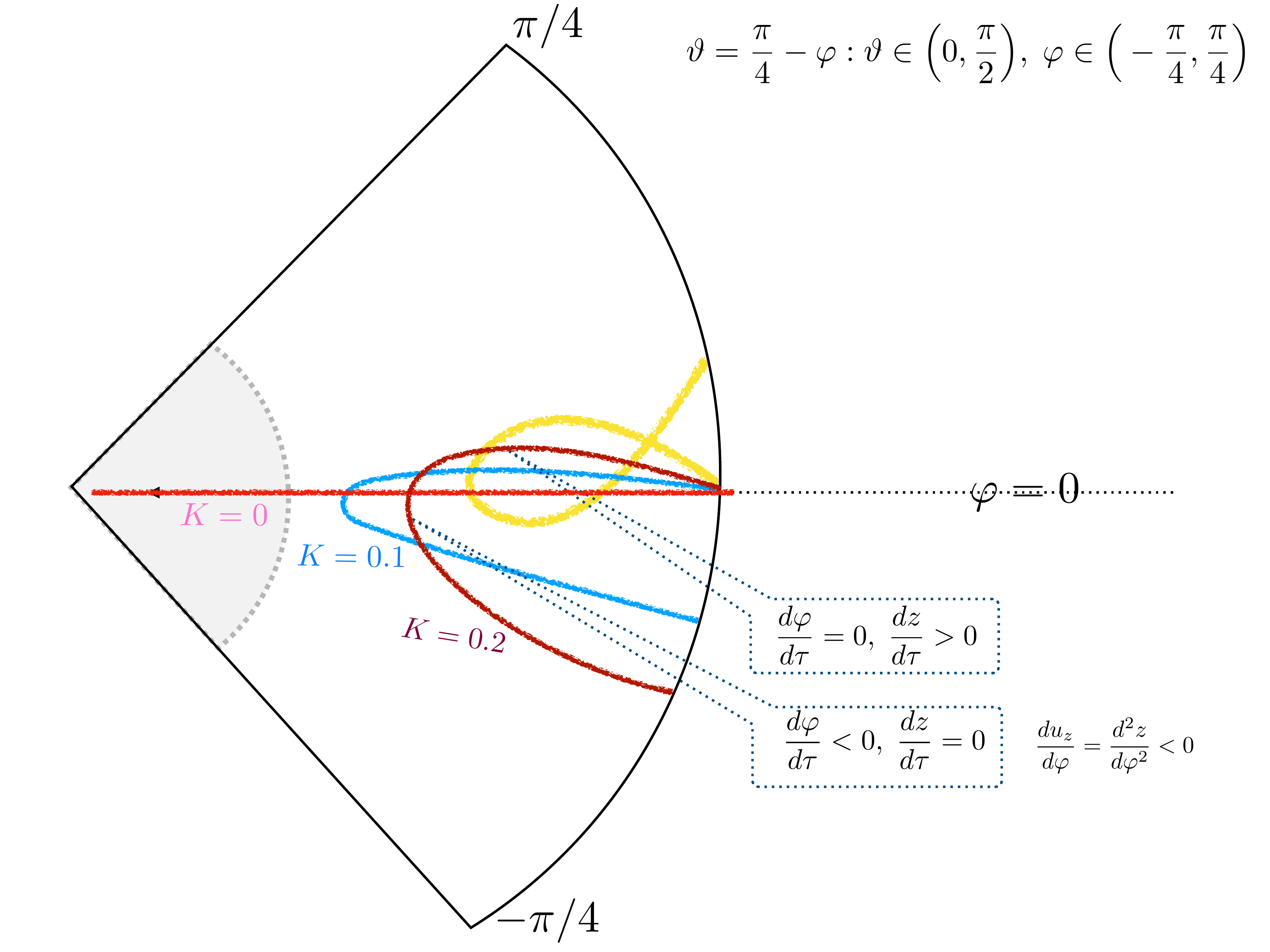}
	\caption{Form of the D5-brane in $(r,\vartheta)$: 
		The area shaded in gray is inside of the horizon}
	\label{fig:iAdSHopf_sketch}
	\end{minipage}
\hspace{0.01\linewidth}
\end{figure}

The whole sketch of the D5-brane is drawn in Figure \ref{fig:iAdSHopf_sketch}.
The metric \eqref{eq:dsindAdS5S5} includes the two cyclic coordinates, $\phi_1$ and $\phi_2$ which are coordinates on the two circles with radii $\sin\vartheta$ and $\cos\vartheta$.
Then this induced metric has the structure which includes the product of these two circles.
The radii of two circles change depending on $\varphi = \pi/4 - \vartheta$.
At the edges $\varphi = \pm\pi/4$ one of these circles shrinks to zero.
We assumed that one side of the brane touches to the boundary at $\varphi=0$ since $\varphi=0$ is only the solution that can penetrate the horizon and we want to follow the changes according to the value of the gauge flux.
Then we found that if the brane has the non zero gauge flux the brane cannot touch the horizon.
The smaller the value of the gauge flux, the closer the D5-brane approaches the horizon.

\section{Interface in the Kerr-AdS black hole spacetime}\label{sec:iKABH}
Here we add the black hole angular momentum to the black hole spacetime considered in the previous section.
Its metric is called Kerr-AdS black hole.

\subsection{Action and Equation of motion}
We set the coordinates on the AdS subspace and $S^5$ subspace as follows:
\begin{align}
\text{AdS}_5&:\qquad
	t, \phi_1, \phi_2, r, \vartheta;
&&	\vartheta\in[0,\pi/2],\qquad
	\phi_1,\phi_2\in[0,2\pi),\nonumber\\
S^5&:\qquad
	\theta, \phi, \varphi_1, \varphi_2, \varphi_3,
&&	\theta,\phi,\varphi_1,\varphi_2\in[0,\pi],\qquad
	\varphi_3\in[0,2\pi).
\end{align}
The worldvolume of the D5-brane extends over the subspace spanned by $(t,r,\vartheta,\phi_1,\phi_2,\theta,\phi)$, where it forms a curve in $(r,\vartheta)$ plane.
Then the embedding of the D5-brane is expressed as a function $\vartheta$ of $r$
\begin{equation}\label{eq:iKA_theta_r}
\vartheta = \vartheta(r).
\end{equation}
We take the ansatz for the gauge flux as before
\begin{equation}\label{eq:iKA_gauge_flux}
\mathcal F = -\kappa d\theta\wedge\sin\theta d\phi.
\end{equation}
The compact part is induced to
$
ds_{S^5}^2 = d\theta^2 + \sin^2\theta d\phi^2
$
on the D5-brane.
The non-compact part is the 5-dimensional Kerr-AdS black hole \cite{Cai:2016xho}.
Its metric is 
\begin{align}
ds_\text{KA5}^2
&= -\frac{\Delta_r}{\rho^2}
  \Big(dt-\frac{a\sin^2\theta}{\Xi_a}d\phi_1 - \frac{b\cos^2\theta}{\Xi_b}d\phi_2\Big)^2\nonumber\\
& + \frac{\Delta_\theta\sin^2\theta}{\rho^2}
  \Big(adt-\frac{r^2+a^2}{\Xi_a}d\phi_1\Big)^2
 + \frac{\Delta_\theta\cos^2\theta}{\rho^2}
  \Big(bdt-\frac{r^2+b^2}{\Xi_b}d\phi_2\Big)^2\nonumber\\
& + \frac{\rho^2}{\Delta_r}dr^2 + \frac{\rho^2}{\Delta_\theta}d\theta^2
 + \frac{1+r^2}{r^2\rho^2}
   \Big(abdt 
  - \frac{b(r^2+a^2)\sin^2\theta}{\Xi_a}d\phi_1
  - \frac{a(r^2+b^2)\cos^2\theta}{\Xi_b}d\phi_2\Big)^2.
\end{align}
In the above
\begin{align*}
&
\rho^2(r) = r^2 + a^2\cos^2\theta + b^2\sin^2\theta,\\
&
\Delta_r(r) = \frac1{r^2}(r^2+a^2)(r^2+b^2)(r^2+1) - 2m,\qquad
\Delta_\theta(\theta) = 1 - a^2\cos^2\theta - b^2\sin^2\theta,\\
&
\Xi_a = 1-a^2,\qquad
\Xi_b = 1-b^2.
\end{align*}
Since the rotations around the two coordinetes $\phi_1$ and $\phi_2$ are equivalent, we focus on the case where the black hole rotates either of two axis ($a\neq0, b=0$).
Then the metric simplifies to 
\begin{align}\label{eq:interfaceKA5}
ds_\text{KA5}^2
&= -\frac{\Delta_{ra}}{\rho_a^2}
  \Big(dt-\frac{a\sin^2\vartheta}{\Xi_a}d\phi_1\Big)^2
 + \frac{\Delta_{\vartheta a}\sin^2\vartheta}{\rho_a^2}
  \Big(adt-\frac{r^2+a^2}{\Xi_a}d\phi_1\Big)^2
 + \frac{\Delta_{\vartheta a}\cos^2\vartheta}{\rho_a^2}r^4d\phi_2^2\nonumber\\
&\qquad
 + \frac{\rho_a^2}{\Delta_{ra}}dr^2 + \frac{\rho_a^2}{\Delta_{\vartheta a}}d\vartheta^2
 + \frac{1+r^2}{\rho_a^2}a^2r^2\cos^4\vartheta d\phi_2^2\nonumber\\
&= -\frac{\Delta_{ra}-\Delta_{\vartheta a}a^2\sin^2\vartheta}{\rho_a^2}dt^2
 + \frac{-\Delta_{ra}a^2\sin^2\vartheta + \Delta_{\vartheta a}(r^2+a^2)^2}{\rho_a^2\Xi_a^2}\sin^2\vartheta d\phi_1^2\nonumber\\
&\qquad
 + 2\frac{\Delta_{ra} - (r^2+a^2)\Delta_{\vartheta a}}{\rho_a^2\Xi_a}a\sin^2\vartheta dtd\phi_1
 + r^2\cos^2\vartheta d\phi_2^2
 + \frac{\rho_a^2}{\Delta_{ra}}dr^2 + \frac{\rho_a^2}{\Delta_{\vartheta a}}d\vartheta^2,
\end{align}
where
\begin{align}
\rho_a^2 
&= r^2 + a^2\cos^2\vartheta,\;\;
\Delta_{ra}
= (r^2+a^2)(r^2+1) - 2m,\;\;
\Delta_{\vartheta a}
= 1 - a^2\cos^2\vartheta,\;\;
\Xi_a = 1-a^2.
\end{align}
The horizon is located at, by coordinate $z:=1/r$,
\begin{align}
\Delta_{ra} = 0 \Rightarrow
(2m-a^2)z^4 - (1+a^2)z^2 - 1 = 0,\nonumber\\
z_\text{h}^2 = \frac{1+a^2+\sqrt{(1+a^2)^2+8m-4a^2}}{4m-2a^2}.
\end{align}

The Ramond-Ramond 4-form is determined by
$
4d\text{vol}[\text{AdS}_5] = dC_4.
$
Let us find the volume form of the AdS part.
The above metric \eqref{eq:interfaceKA5} can be written as 
\begin{equation}
ds_\text{KA5}^2
= -\frac{\Delta_{ra}}{\rho_a^2}dt'^2
  + \frac{\Delta_{\vartheta a}\sin^2\vartheta}{\rho_a^2}d\phi_1'^2
  + \text{diag}(r,\vartheta,\phi_2),\;\;
 \begin{bmatrix}
dt'\\ d\phi_1'
\end{bmatrix}
=
M\begin{bmatrix}
dt\\ d\phi_1
\end{bmatrix},
\end{equation}
where
\begin{align*}
M &:=
\begin{bmatrix}
1& -a\sin^2\vartheta/\Xi_a\\
a& -(r^2+a^2)/\Xi_a
\end{bmatrix},\\
\det M
&= -\frac{r^2+a^2(1-\sin^2\vartheta)}{\Xi_a}
= -\frac{r^2+a^2\cos^2\vartheta}{\Xi_a}
= -\frac{\rho_a^2}{\Xi_a}.
\end{align*}
Then the determinant of the metric for $t$ and $\phi_1$ part is
\begin{equation}
\det G = \det(M^TG'M)
= (\det M)^2\det G'
= - \frac{\Delta_{ra}\Delta_{\vartheta a}\sin^2\vartheta}{\rho_a^4}\frac{\rho_a^4}{\Xi_a^2}
= - \frac{\Delta_{ra}\Delta_{\vartheta a}\sin^2\vartheta}{\Xi_a^2},
\end{equation}
and the total factor in AdS is 
\begin{equation}
 - \frac{\Delta_{ra}\Delta_{\vartheta a}}{\Xi_a^2}r^2\sin^2\vartheta\cos^2\vartheta
\frac{\rho_a^2}{\Delta_{ra}}\frac{\rho_a^2}{\Delta_{\vartheta a}}
= - \frac{\rho_a^4}{\Xi_a^2}r^2\sin^2\vartheta\cos^2\vartheta.
\end{equation}
The volume form of the AdS part is
\begin{equation}
d\text{vol}_\text{AdS} 
=  \frac{\rho_a^2r\sin\vartheta\cos\vartheta}{\Xi_a}
	dtdrd\vartheta d\phi_1d\phi_2.
\end{equation}
Then the Ramond-Ramond 4-form is
\begin{equation}
C_4 
= - \frac{\rho_a^4}{\Xi_a}\sin\vartheta\cos\vartheta dtd\vartheta d\phi_1d\phi_2.
\end{equation}

From the assumption for the embedding \eqref{eq:iKA_theta_r} and the gauge flux \eqref{eq:iKA_gauge_flux}, the sum of the induced metric and the gauge flux is
\begin{align*}\label{eq:interfaceKAGF}
& G_\text{ind} + \mathcal F =\\
&
\begin{bmatrix}
-\frac{\Delta_{ra}-\Delta_{\vartheta a}a^2\sin^2\vartheta}{\rho_a^2}& & 
  \frac{\Delta_{ra}-\Delta_{\vartheta a}(r^2+a^2)}{\rho_a^2\Xi_a}a\sin^2\vartheta& & &\\
 & \frac{\rho_a^2}{\Delta_{ra}} + \frac{\rho_a^2}{\Delta_{\vartheta a}}\vartheta'^2& & & &\\
\frac{\Delta_{ra}-\Delta_{\vartheta a}(r^2+a^2)}{\rho_a^2\Xi_a}a\sin^2\vartheta& & 
  \frac{-\Delta_{ra}a^2\sin^2\vartheta + \Delta_{\vartheta a}(r^2+a^2)^2}{\rho_a^2\Xi_a^2}\sin^2\vartheta& & &\\
 & & & r^2\cos^2\vartheta& &\\
 & & & & 1& -\kappa\sin\theta\\
 & & & & \kappa\sin\theta& \sin^2\theta
\end{bmatrix}.
\end{align*}
Its determinant is
\begin{align*}
\sqrt{-\det(G_\text{ind} + \mathcal F)}
&= \sqrt{1+\kappa^2}\; \cos\vartheta\sin\vartheta
\frac{r\rho_a}{\Xi_a}\sqrt{\Delta_{ra}\vartheta'^2+\Delta_{\vartheta a}}.
\end{align*}
In the above, since $\vartheta\in[0,\pi/2]$, then $\cos\vartheta \geq 0$.
The D5-brane action consists of the following.
\begin{align}
S_\text{D5} &= S_\text{DBI} + S_\text{WZ},\\
S_\text{DBI}
&= -T_5\int\sqrt{-\det(G_\text{ind} + \mathcal F)},
\nonumber\\
S_\text{WZ}
&= T_5\int\mathcal F\wedge C_4.
\nonumber
\end{align}
We change the variables $z = 1/r$, $\varphi = \pi/4 - \vartheta$.
Introducing a parameter $\tau$, $(z(\tau),\varphi(\tau))$,
\begin{subequations}
\begin{align}
\sqrt{-\det(G_\text{ind}+\mathcal F)} 
&= \sqrt{1+\kappa^2}d\tau\cos(2\varphi)\frac{r\rho_a}{2\Xi_a}
	\sqrt{\Delta_{ra}\dot\varphi^2 + \Delta_{\varphi a}\dot r^2},\\
\mathcal F\wedge C_4
&= \frac{\rho_a^4}{2\Xi_a}d\varphi\cos(2\varphi).
\end{align}
\end{subequations}
The action is
\begin{equation}\label{eq:interfaceKAaction}
S_\text{D5} 
= -T_5(2\pi)^2\text{vol}[S^2]T\int d\tau\mathcal L,
\end{equation}
where the Lagrangian is  
\begin{align}
\mathcal L 
&= \frac{\sqrt{1+\kappa^2}}{2\Xi_a}\cos(2\varphi)\Big(
\frac{\rho_a}{z^3}\sqrt{\Delta_{\varphi a}\dot z^2+z^4\Delta_{ra}\dot\varphi^2}
- K\rho_a^4\dot\varphi\Big)\nonumber\\
&= \frac{\sqrt{1+\kappa^2}}{2\Xi_a}\frac{\cos(2\varphi)}{z^4}\Big(
p_a\sqrt{\Delta_{\varphi a}\dot z^2+\Delta_{za}\dot\varphi^2}
- Kp_a^4\dot\varphi\Big).
\end{align}
In the second line of the above we defined 
\begin{subequations}
\begin{align}
&\rho_a^2 = \frac1{z^2} + \frac{a^2}2(1+\sin(2\varphi))
=: \frac{p_a^2}{z^2},\;\;
\Delta_{ra} 
= \Big(\frac1{z^2}+a^2\Big)\Big(\frac1{z^2}+1\Big) - 2m
=: \frac{\Delta_{za}}{z^4},\\
&\Delta_{\varphi a}
:= 1-\frac{a^2}2(1+\sin(2\varphi)).
\end{align}
\end{subequations}
We choose the gauge where
$\Delta_{\varphi a}\dot z^2 + \Delta_{za}\dot\varphi^2 = 1$ 
to fix the gauge degrees of freedom.

The equations of motion are
\begin{subequations}
\begin{align}
z:\;\;
&\frac{d}{d\tau}\Big(\frac{\cos(2\varphi)}{z^4}
p_a\Delta_{\varphi a}\dot z\Big)
+ \frac{4\cos(2\varphi)}{z^5}(p_a - Kp_a^4\dot\varphi)\nonumber\\
&\hspace{5cm}
- \frac{\cos(2\varphi)}{z^4}(
  \partial_zp_a + \frac12p_a\partial_z\Delta_{za}\dot\varphi^2 
  - 4Kp_a^3\partial_zp_a\dot\varphi
) = 0,\\
\varphi:\;\;
&\frac{d}{d\tau}\Big(\frac{\cos(2\varphi)}{z^4}p_a\Delta_{za}\dot\varphi\Big)
- \frac{d}{d\tau}\Big(\frac{\cos(2\varphi)}{z^4}Kp_a^4\Big)
+ \frac{2\sin(2\varphi)}{z^4}(p_a - Kp_a^4\dot\varphi)\nonumber\\
&\hspace{5cm}
- \frac{\cos(2\varphi)}{z^4}(
  \partial_\varphi p_a + \frac12p_a\partial_\varphi\Delta_{\varphi a}\dot z^2 
  - 4Kp_a^3\partial_\varphi p_a\dot\varphi
) = 0.
\end{align}
\end{subequations}
Let us define variables $u_z := dz/d\tau$ and $u_\varphi := d\varphi/d\tau$. 
Then the equations are
\begin{subequations}\label{eq:KABH_eom}
\begin{align}
\frac{dz}{d\tau} &= u_z,\\
\frac{d\varphi}{d\tau} &= u_\varphi,\\
\frac{du_z}{d\tau} &= 
u_z\Big(\frac{4u_z}z - \frac{\dot p_a}{p_a} - \frac{\dot\Delta_{\varphi a}}{\Delta_{\varphi a}} + 2u_\varphi\tan(2\varphi)\Big)\nonumber\\
&\qquad
+ \Delta_{\varphi a}^{-1}\Big(
  \frac{\partial_zp_a}{p_a}(1-4Kp_a^3u_\varphi)
  + \frac12\partial_z\Delta_{za}u_\varphi^2
\Big)
- \frac4z\Delta_{\varphi a}^{-1}(1 - Kp_a^3u_\varphi),\\
\frac{du_\varphi}{d\tau}
&= u_\varphi\Big(\frac{4u_z}z - \frac{\dot p_a}{p_a} - \frac{\dot\Delta_{za}}{\Delta_{za}}    + 2u_\varphi\tan(2\varphi)\Big)
 + \Big(\frac{\cos(2\varphi)}{z^4}p_a\Delta_{za}\Big)^{-1}
    \frac{d}{d\tau}\Big(\frac{\cos(2\varphi)}{z^4}Kp_a^4\Big)\nonumber\\
&\qquad
+ \Delta_{za}^{-1}\Big(
  \frac{\partial_\varphi p_a}{p_a}(1-4Kp_a^3u_\varphi)
  + \frac12\partial_\varphi\Delta_{\varphi a}u_z^2\Big)
  - 2\tan(2\varphi)\Delta_{za}^{-1}(1-Kp_a^3u_\varphi).
\end{align}
\end{subequations}

Let us confirm the validity of the constraint. 
By substituting these equations of motion,
\begin{align*}
&\frac{d}{d\tau}(\Delta_{\varphi a}u_z^2 + \Delta_{za}u_\varphi^2)
= 2\Delta_{\varphi a}u_z\dot u_z + 2\Delta_{za}u_\varphi\dot u_\varphi
 + \Delta_{\varphi a}'(\varphi)u_z^2u_\varphi + \Delta_{za}'(z)u_zu_\varphi^2\\
&= -2\Delta_{\varphi a}u_z^2\frac{d}{d\tau}\log\Big(\frac{\cos(2\varphi)}{z^4}p_a\Delta_{\varphi a}\Big)
 - \frac{8u_z}z(1-Kp_a^3u_\varphi)
 + 2u_z\Big(\frac{\partial_zp_a}{p_a}+\frac12\partial_z\Delta_{za}u_\varphi^2-4Kp_a^2\partial_zp_au_\varphi\Big)\\
&\qquad
 - 2\Delta_{za}u_\varphi^2\frac{d}{d\tau}\log\Big(\frac{\cos(2\varphi)}{z^4}p_a\Delta_{za}\Big)
 - 4u_\varphi\tan(2\varphi)(1-Kp_a^2u_\varphi)\\
&\qquad
 + 2u_\varphi\Big(\frac{\partial_\varphi p_a}{p_a}+\frac12\partial_\varphi\Delta_{\varphi a}u_z^2-4Kp_a^2\partial_\varphi p_au_\varphi\Big)
 + 2Kp_a^3u_\varphi\frac{d}{d\tau}\log\Big(\frac{\cos(2\varphi)}{z^4}p_a^4\Big)
 + \Delta_{\varphi a}'u_z^2u_\varphi 
 + \Delta_{za}'u_zu_\varphi^2\\
&= -2(1-Kp_a^3u_\varphi)(-2\tan(2\varphi) - \frac{4u_z}z)
 - \frac{8u_z}z(1-Kp_a^3u_\varphi)
 - 4u_\varphi\tan(2\varphi)(1-Kp_a^2u_\varphi)
= 0.
\end{align*}
Then this constraint is valid.

\paragraph{Initial condition}
As in the previous section, we use the approximation $\varphi = \arcsin(\kappa z)$ near the boundary.
Taking account of the constraint, $\Delta_{\varphi a}u_z^2 + \Delta_{za}u_\varphi^2 = 1$, for small $z_0$ 
\begin{subequations}\label{eq:KABH_initial}
\begin{align}
\varphi_0
&= \arcsin(\kappa z_0),\\
u_z(\tau_0) 
&= \sqrt\frac{1-\kappa^2z^2}{\Delta_{\varphi a}(1-\kappa^2z^2) + \Delta_{za}\kappa^2},\\
u_\varphi(\tau_0) 
&= \frac{\kappa}{\sqrt{\Delta_{\varphi a}(1-\kappa^2z^2) + \Delta_{za}\kappa^2}}.
\end{align}
\end{subequations}

\subsection{Interface solution}\label{subsec:iKA_sol}
We perform the numerical calculation to solve the equations of motion \eqref{eq:KABH_eom} under the initial condition \eqref{eq:KABH_initial}.
The whole structure of bulk spacetime is the same as figure \ref{fig:iAdSHopf_sketch}.
The result is shown in the following two figures.

\begin{figure}[h]
	\begin{minipage}[t]{0.5\linewidth}
	\includegraphics[width=\linewidth]{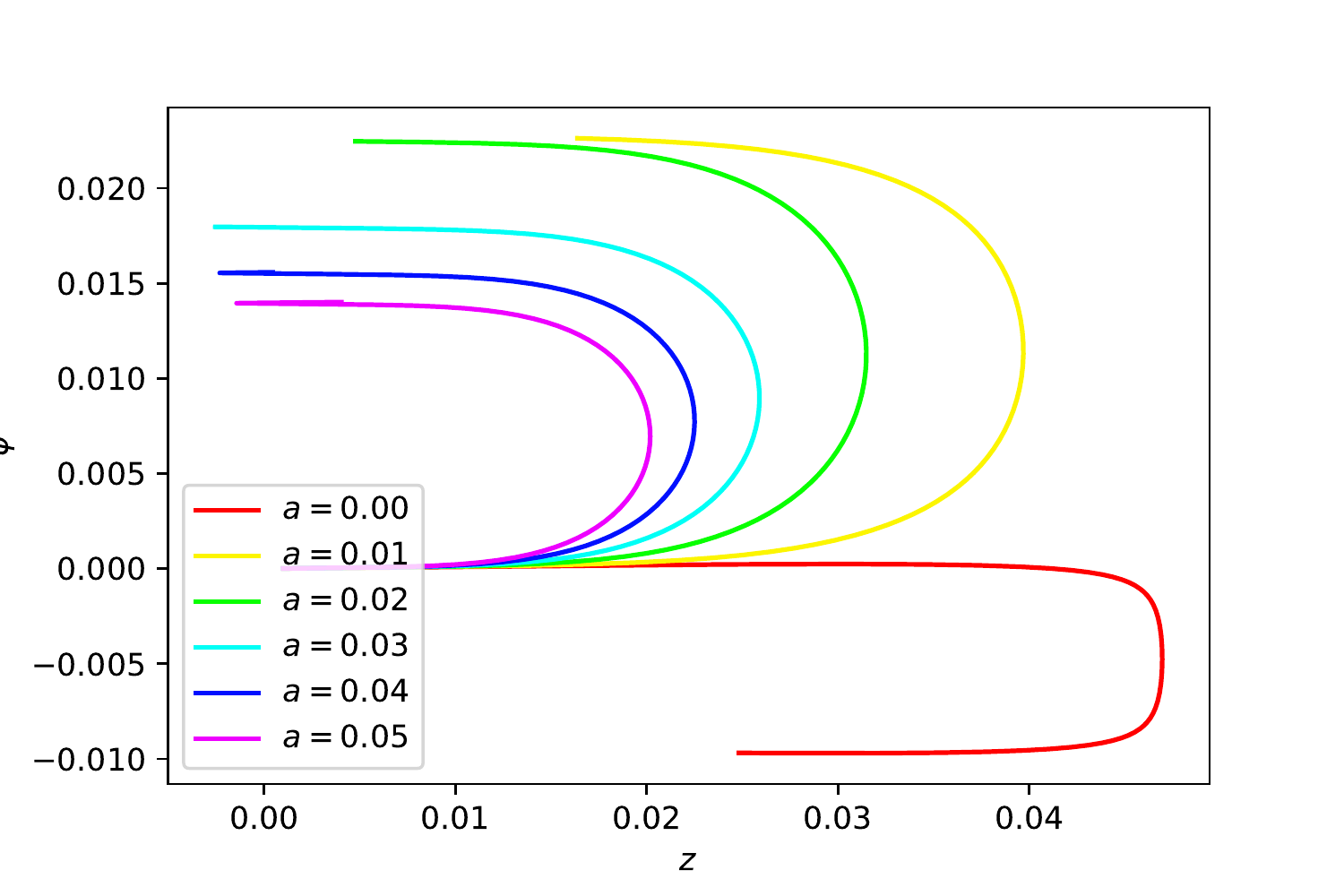}
	\caption{Angular momentum $a$ dependence (1): $m$ = $10^5$, $K=0.01$}
	\label{fig:iKA_a1}
	\end{minipage}
\hspace{0.01\linewidth}
	\begin{minipage}[t]{0.5\linewidth}
	\includegraphics[width=\linewidth]{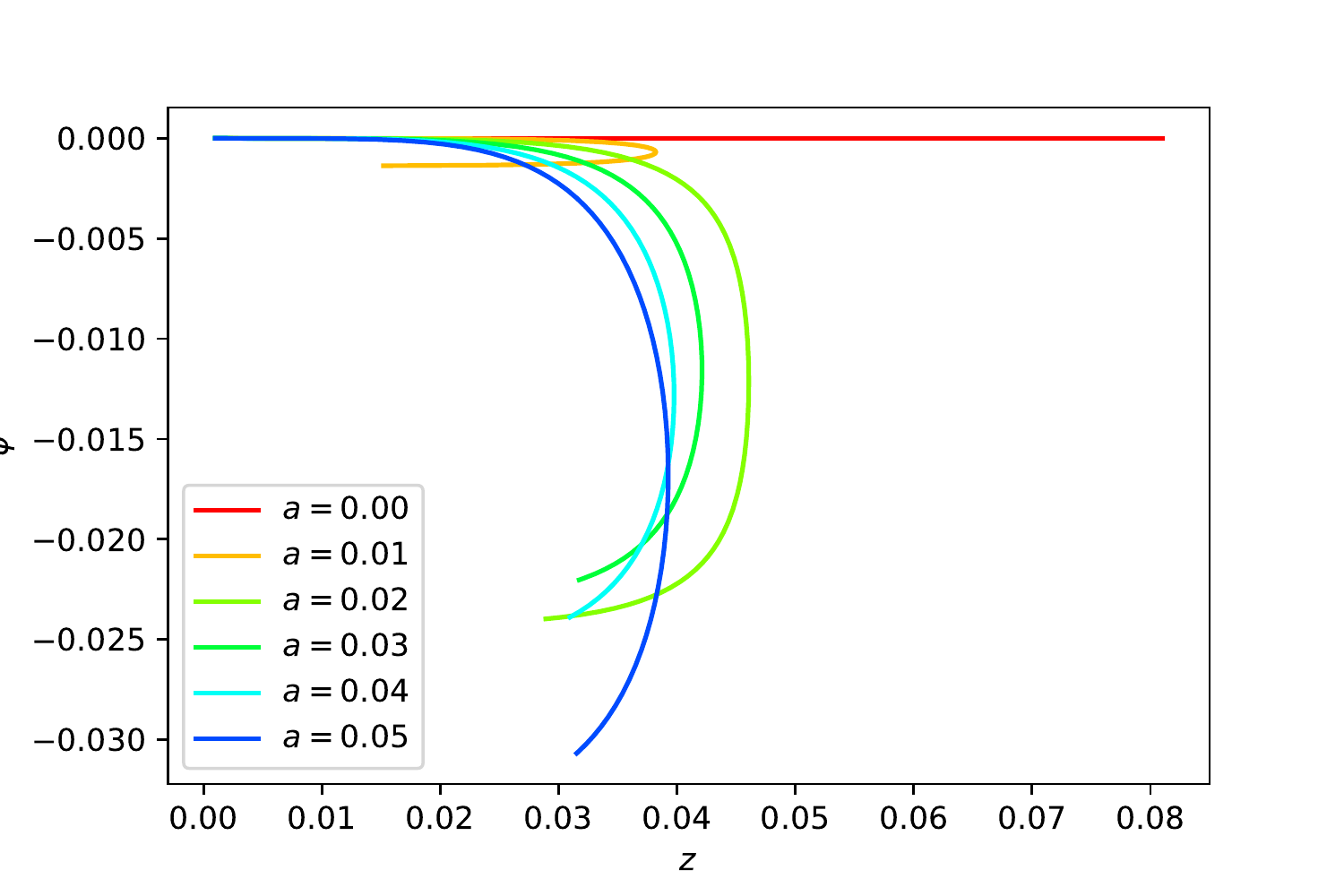}
	\caption{Angular momentum $a$ dependence (3): $m$ = $10^5$, $K=0$}
	\label{fig:iKA_a3}
	\end{minipage}
\end{figure}

Figure \ref{fig:iKA_a1} shows the angular momentum dependence. 
This tells us that the D5-brane cannot enter the horizon in the same to the AdS Schwarzschild black holes.
The black hole mass is fixed to $m = 10^5$ and the the strength of the flux is fixed to $K=0.01$.
The location of the horizon is $z_\text{h}\approx 0.047$ for $a = 0$.


Figure \ref{fig:iKA_a3} shows the $a$ dependence for $K=0$.
We see that the D5-brane can enter the horizon only for $a=0$.
Even if the gauge flux is zero $K=0$, the D5-brane avoids the horizon for the black holes which have the non zero angular momentum.
For example the horizon is located at $z_\text{h} \approx 0.047$ for $a=0.05$.

\section{Summary and discussion}\label{Sec:Discussion}
In this paper we treat the interface which is a co-dimension one object realized by the D5-brane in IIB string theory.
We obtained the interface solution in the AdS black hole spacetime by the numerical calculation.
The boundary condition is given in the same way for the flat AdS$_5\times S^5$ case imposed in \cite{Nagasaki:2011ue}, that is, the degree of the slope at the boundary changes according to the gauge flux.
The D5-brane is bent inside of spacetime by the effect of the black hole.
The D5-brane penetrates the horizon only for the case $K=0$ and $a=0$.
We found in \cite{Nagasaki:2018dyz} that this probe brane can enter the horizon only for the case $K=0$ and in this paper that it cannot enter the horizon if the black holes have the angular momentum, either.

\begin{figure}[h]
	\includegraphics[width=\linewidth]{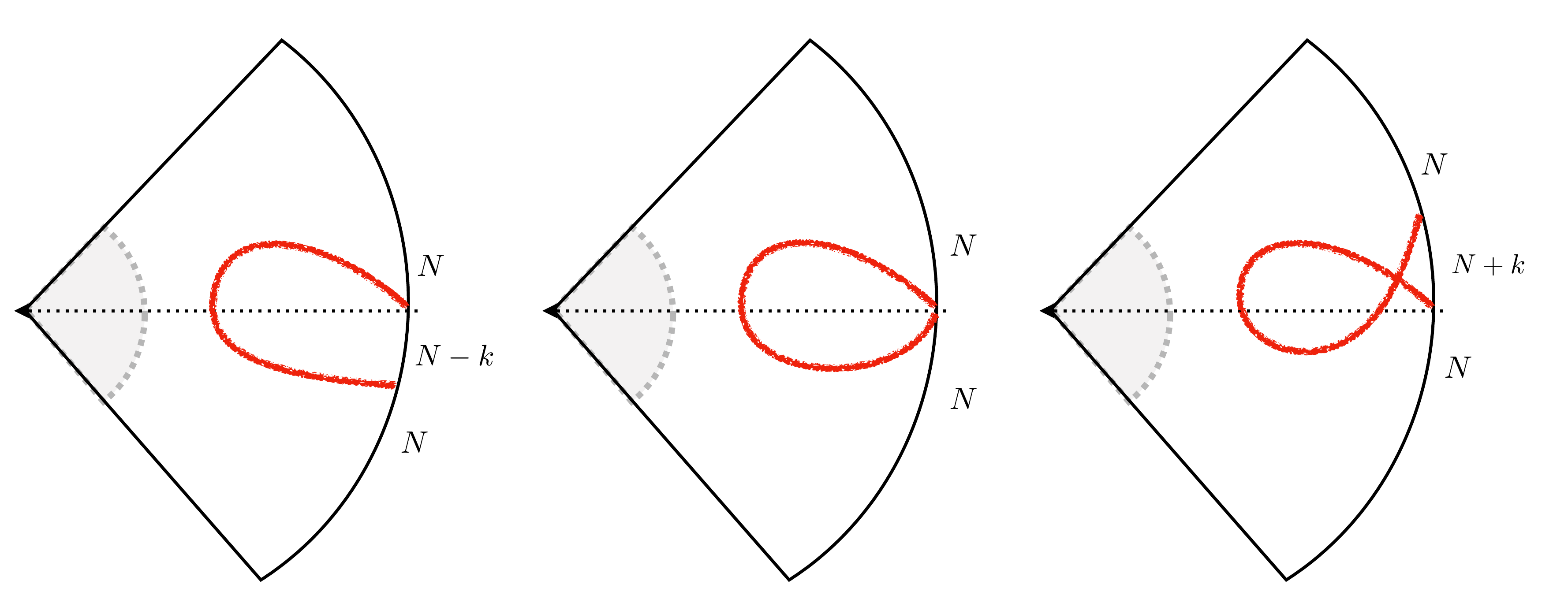}
	\caption{A sketch of the transition according to the gauge flux}
	\label{fig:iAdSBH_hopf_transition}
\end{figure}

For boundary theories, we can also have an interesting conclusion.
By the gauge/gravity correspondence, the bulk has super Yang-Mills theories with gauge groups SU($N$) and SU($N-k$) on the boundary.
Here the difference $k$ is related to the parameter $\kappa$ as we can check by integrating the gauge flux on the D5-brane  \cite{Nagasaki:2011ue}
\begin{equation}
k = -\frac{T_5}{T_3}\int\mathcal F = \frac{\kappa}{\pi\alpha'}.
\end{equation}
The edge of the probe D5-brane touching the boundary corresponds the interface which separates these two gauge theories. 
As we saw in this paper, the static probe D5-brane without the gauge flux is the only solution which  penetrates the horizon.
Then in this case, there is the interface located at $\varphi = 0$ and there are two gauge theories over $-\pi/4\leq\varphi\leq0$ and $0\leq\varphi\leq\pi/4$ (see Figure \ref{fig:iAdSHopf_sketch}).
If the gauge flux is non zero, the probe brane bends and touches the boundary at two positions.
Then in this case there are two interfaces and there are three gauge theories at the boundary.
The difference of the gauge groups between both sides of the interface is given by $k$.
We see the transition from small $k$ to large $k$ (see Figure \ref{fig:iAdSBH_hopf_transition}).
Suppose the gauge group of the below side of the interface at $\varphi=0$ is lower than the upper side (left of Figure \ref{fig:iAdSBH_hopf_transition}). 
There are two interfaces in this case and they separate the boundary into three gauge theories with gauge groups SU($N$), SU($N-k$) and SU($N$).
As the strength of the gauge flux increases, the D5-brane bends more tightly. 
At the critical point, the two interfaces coincides and degenerate to an interface with the same gauge group SU($N$) on the both sides (middle of Figure \ref{fig:iAdSBH_hopf_transition}).
If the strength of the gauge flux exceeds the critical point, then a self-intersection occurs (right of Figure \ref{fig:iAdSBH_hopf_transition}).
For the interface at $\varphi=0$, the gauge group is lower in the below side as we can see in the left figure.
Then a gauge theory with SU($N+k$) occurs on the upper side of this interface.
Then, in the resulting situation, there are three gauge theories with gauge groups SU($N$), SU($N+k$) and SU($N$).

In this paper we considered the case the black holes has the angular momentum but the probe brane is static.
Considering a moving defect \cite{Janiszewski:2011ue} is important to test CA conjecture.
As stated in \cite{Hubeny:2012ry}, the extremal surfaces do not penetrate the horizon in a static case, whereas they penetrate the horizon in the time dependent case.
Then there is a possibility that the probe brane can also penetrate the horizon if we consider a dynamical interface or other non-local dynamical operators.
Since the growth of the action is calculated in the inside of the horizon, it is important to study  non-local operators which affects this region. 
It will be important theme to study CA conjecture and black hole complexity.

\section*{Acknowledgement}
I would like to thank Satoshi Yamaguchi, UESTC members and people discussing at the 74th meeting of the Physical Society of Japan.


\providecommand{\href}[2]{#2}\begingroup\raggedright\endgroup

\end{document}